\documentclass[11pt]{article}
\usepackage{graphicx} % Required for inserting images
\usepackage[table]{xcolor}
\usepackage{makecell}
\usepackage{multirow}
\usepackage[dvipsnames]{xcolor}
\usepackage{hyperref}
\usepackage{amsmath}
\usepackage{booktabs}
\hypersetup{
    colorlinks,
    linkcolor={red!50!black},
    citecolor={blue!50!black},
    urlcolor={blue!80!black}
}
\usepackage[margin = 1in]{geometry}
\usepackage{placeins}

\title{Census Dual Graphs: Properties and Random Graph Models}

\author{Sara Anderson\thanks{\texttt{sranders15@gmail.com}, 
Claremont Graduate University, Claremont, CA}, 
Sarah Cannon\thanks{\texttt{scannon@cmc.edu}, 
Claremont McKenna College, Claremont, CA}, %; corresponding author; ORCID 0000-0001-6510-4669}, 
Brooke Feinberg\thanks{\texttt{bfeinber@usc.edu}, 
Universiy of Southern California, Los Angeles, CA}, Anne Friedman\thanks{\texttt{akfriedman13@gmail.com}, 
Scripps College, Claremont, CA}}

\date{}

\begin{document}

\maketitle

\begin{abstract}

In the computational study of political redistricting, feasibility necessitates the use of a discretization of regions such as states, counties, and towns. In nearly all cases, researchers use a {\it dual graph}, whose vertices represent small geographic units (such as census blocks or voting precincts) with edges for geographic adjacency. A political districting plan is a partition of this graph into connected subgraphs that satisfy certain additional properties, such as connectedness, compactness, and equal population. 

Though dual graphs underlie nearly all computational studies of political redistricting, little is known about their properties. This is a unique graph class that has been described colloquially as `nearly planar, nearly triangulated,' but thus far there has been a lack of evidence to support this description.  In this paper we study dual graphs for counties, census tracts, and census block groups across the United States in order to understand and characterize this graph class. We also consider several random graph models (most based on randomly perturbing grids or Delauney triangulations of random point sets), and determine which most closely resemble dual graphs under key metrics.  This work lays an initial foundation for understanding and modeling the properties of dual graphs; this will provide invaluable insight to researchers developing algorithms using them to understand, assess, and quantify the properties of political districting plans. 

\vspace{5mm}

\vspace{5mm}
\noindent {\bf Acknowledgements: } S. Cannon is supported in part by National Science Foundation grants CCF-2104795 and CCF-2443221. 

\end{abstract}

\newpage

\section{Introduction}

Across the United States and the world, regions are divided into districts for the purposes of electing representatives.  %How these districts are drawn can have a huge effect on who is ultimately elected.  
  Redrawing district boundaries can have significant implications for electoral outcomes, shaping which candidates are elected and which political parties gain power. As a result, redistricting is often a highly contentious process. Drawing districts so as to give an electoral advantage to a certain individual, group, or party is known as {\it gerrymandering}.

A variety of computational tools and approaches have been developed to detect and quantify gerrymandering, as well as to study political districting and the space of possible districting plans more generally.  For example, sampling algorithms can create a variety of  districting plans satisfying criteria such as equal population across districts, contiguity, and respect for geographic boundaries \cite{DukeReCom1,  DeFord2021Recombination, smc, ImaiFlip}. This can be used to understand what a `typical' districting plan looks like in the absence of other motivations, and thus
 when a particular districting plan is an outlier (that is, it may be gerrymandered) \cite{compRedistVRA, va-criteria, georgia}.  Other uses of sampling algorithms in the context of political redistricting include assessing a map-drawer's stated motivation or understanding the possible effects of rule changes~\cite{ chenstephanopoulos, cohen2021dp, MRL}.

\subsection{Dual Graphs for Geography}\label{sec:dual_graph_geo}

While many people think of political districts as being defined by lines drawn through and around communities, this viewpoint is challenging from a computational point of view: The sheer number of possibilities for these lines make analysis challenging and likely infeasible. Instead, districts can be viewed as being built by aggregating small geographic pieces.  For example, U.S. Congressional districts cannot split census blocks, so one can view a Congressional district as a connected union of census blocks. Districts also rarely split larger units, such as census block groups or voting precincts, so the same perspective is again useful here as well. 

Computational approaches to studying redistricting thus use a discretization of geography into graph structures called dual graphs. In these geographic dual graphs, each vertex corresponds to a geographic unit (such as a census block or a voting precinct), and an edge connects two vertices if their corresponding regions share a border \cite{duchin2023discrete}. A political district is a connected subgraph of a dual graph, an a political districting plan is a partition of the dual graph into connected districts with approximately equal population (which we sometimes refer to as being {\it balanced}). The use of dual graphs allow researchers to perform analyses %, such as testing for gerrymandering, and generate alternative districting plans using algorithmic methods 
that would not otherwise be computationally feasible. 

Little is known about these dual graphs. %, which are essential in computational redistricting. 
%Constructing geographic dual graphs requires access to shapefiles and geospatial data, which can be difficult to obtain or work with, especially for large-scale simulations or theoretical analyses. 
Without a better understanding of the structural properties of these graphs, it is difficult to design or evaluate algorithms effectively. A better understanding of dual graphs and their properties can guide future algorithmic exploration by clarifying the graph class on which algorithms need to be correct and efficient.  %In addition, a better understanding of the structure of real-world dual graphs can help researchers identify which algorithms are most effective for working with them.

Constructing geographic dual graphs requires access to shapefiles and geospatial data, which can be difficult to obtain or work with, especially for large-scale simulations or theoretical analyses.  This also motivates the need for a random graph model that can approximate the key structural properties of dual graphs without relying on actual map data. A model like this could serve as a testbed for redistricting algorithms and support theoretical investigations into fairness, randomness, and bias in districting plans. If we can identify common structural features, such as degree distribution and connectivity, we can develop models that approximate them without requiring the underlying geographic datasets.

\subsection{Motivation and Related Work}

%\todo{survey the sampling methods, including why spanning trees and splittability are relevant, and that some of the works has made used grids as a simplified model - but how realistic are they really?  }

This section outlines some previous work in computational redistricting. As we are motivated by improving state-of-the-art sampling algorithms, we focus on the previous work most relevant to sampling, though of course there is a large body of work taking other computational approaches to redistricting, including but certainly not limited to optimization~\cite{ValidiBuchanan2022opt}, topological data analysis~\cite{Duchin2022homol}, and optimal transport~\cite{Abrishami2020opttrans}. 

%This includes what assumptions about dual graphs have been made in order to enable theoretical guarantees in various settings. 

\subsubsection{Sampling Algorithms}

In order to combat gerrymandering, some states and researchers have turned towards computational tools to detect and quantify gerrymandering. 
%, both to generate potential districting plans and to evaluate whether existing plans are fair or intentionally biased \cite{dukeComp}. 
One such tool is the Markov Chain Monte Carlo method (MCMC) which generates random samples from a large set of possibilities. In the context of redistricting, MCMC is used to create and analyze many different districting plans that all follow the same basic rules—for example, equal population across districts, contiguity, and respect for geographic boundaries \cite{deford2021mcmc}.  Because the number of valid ways to draw district maps is enormous, it’s impossible to look at every single one. Instead, MCMC provides a way to walk through the space of possible plans, moving from one plan to another by making small changes. Over time, this random walk produces a large, diverse collection of plans that can be analyzed to understand what's typical or unusual in the context of a state or region's geographic distribution of voters.  

All of the sampling methods we will discuss take as input a dual graph representing the geography of the region to be divided into districts. 
Recall a districting plan with $k$ districts is a partition of this dual graph into $k$ connected, approximately-population-balanced subgraphs.  Initial random walks on districting plans were {\it flip} Markov chains, where in each step a random vertex was moved from one district to another~\cite{ImaiFlip, DukeNC}. However, attempting to use such Markov chains for representative sampling is challenging due to needing to run for an extremely large number of steps, though they have been used in other ways to detect gerrymandering~\cite{CFP, CFMP}.

%Random sampling of political districting plans has become an important tool to help understand the space of possible  plans, beginning with~\cite{chen-rodden-unintentional, chen-rodden-thicket}.  If a current or proposed districting plan is an outlier with respect to a collection of randomly sampled districting plans (called an {\it ensemble} of plans), this is evidence it may be~gerrymandered.
%A given districting plan can be compared to this ensemble along various statistics -- such as seats won per party, compactness, efficiency gap, and more -- to see if it is an outlier. 
%A variety of methods exist for creating ensembles: random merging and exchanging of precincts~\cite{chen-rodden-unintentional, chen-rodden-thicket}, flip Markov chains~\cite{ImaiFlip, frieze2022subexponential,DukeNC}, recombination Markov chains~\cite{recom}, sequential Monte-Carlo~\cite{ImaiReCom}, and a two-stage method incorporating fairness~\cite{GS21,GGRS22}. 

More recently, the focus has shifted to {\it recombination} Markov chains~\cite{DeFord2021Recombination, revrecom, DukeReCom1, DukeReCom2}. In each step of these Markov chains, two random districts are merged, a random spanning tree of this union is drawn, and this spanning tree is split into two approximately-equal-population pieces, if possible. These chains seem to perform much better than flip chains, appearing to reach a random state (with no noticeable influence from the starting state) much more quickly than flip chains, though rigorous convergence time bounds still largely remain elusive.

Other relevant sampling algorithms include the up-down Markov chain~\cite{charikar2022complexity} and Sequential Monte Carlo~\cite{smc}, both of which also critically rely on the use of spanning trees. The up-down Markov chain maintains a {\it spanning forest}, that is, a spanning tree of each district, and repeatedly adds a random edge (either within a district or between districts) and then removes a random edge producing another spanning forest. While this method will produce random connected, compact districts, they could be of wildly differing sizes or have wildly different populations. Thus, this chain needs to be combined with rejection sampling to ensure the resulting partitions are balanced.

\subsubsection{Spanning Trees and Compactness}

A {\it spanning tree} of a graph is a subset of edges that connects all the vertices without forming any cycles. %; a {\it cycle in a graph is a path that starts and ends at the same vertex without retracing any edge. 
Every connected graph has at least one spanning tree, and many graphs have a large number of distinct spanning trees. Graphs with more edges tend to have more spanning trees, making a graph's spanning tree count (its total number of distinct spanning trees) a useful measure of how well-connected the graph is. If $N_{ST}(G)$ is the number of spanning trees of graph $G$, as this number is extremely large we most frequently consider $\ln(N_{ST}(G))$.  This quantity depends heavily on the number of vertices in $G$, so we also consider $\ln(N_{ST}(G))/|V(G))|$. For both square grids and triangular grids, this quantity approaches a fixed constant as $|V(G)| \rightarrow \infty$~\cite{Wu_1977}, and our experiments show a similar pattern for dual graphs.  For these reasons we refer to this quantity $\ln(N_{ST}(G))/|V(G))|$ as the {\it spanning tree constant} of a dual graph.

Specific to the redistricting application, the number of spanning trees of a subgraph (a district) has also been proposed as a measure of the district's compactness~\cite{revrecom}; compactness of districts is required in nearly all applied settings.  At a high level, compact districts have shorter boundaries, which means fewer edges between districts and more edges within districts, leading to a higher spanning tree count. This relationship between spanning tree counts and compactness has been validated both theoretically~\cite{procacciaTuckerFoltz2021compactness} and empirically~\cite{clelland2021compactness}.  It is thus natural to study dual graphs' connectivity properties by looking at measures related to spanning trees. 

%The number of spanning trees of a graph has been proposed as a measure of the graph's compatness and well-connectedness. 

%If a district is more rectangular, then it would have more spanning trees. On the contrary, a long, snakey district with the same number of vertices would have less spanning trees. Therefore, the more spanning trees a graph has, the more compact the district is, which Figure~\ref{fig:st_eg_compact} illustrates.

%While there are 

%\begin{figure}
%\begin{center}
%\includegraphics[scale=0.3]{imgs/math_background_egs/spanning_tree_compact.drawio.png}
%\caption{The graph in (a) has 8 spanning trees, while the graph in (b) has 3 spanning trees.  Having more spanning trees correlates with spatial notions of compactness relevant to redistricting applications. }
%%\label{fig:st_eg_compact}
%\end{center}
%\end{figure}

Furthermore, most algorithms for sampling political districting plans rely heavily on spanning trees of districts at intermediate steps~\cite{smc, DukeReCom1, DukeReCom2, DeFord2021Recombination, cannon2024sampling, smc, revrecom}. Algorithms using spanning trees are expected to be a main focus of future advancements as well, and so understanding the spanning tree structure of dual graphs is critical.

\subsubsection{Splittability}

A key subtroutine of recombination algorithms~\cite{DeFord2021Recombination, revrecom, DukeReCom1, DukeReCom2} draws a random spanning tree of the union of two districts, and then checks whether it can be split into two approximately-equally-sized pieces to form two new districts. The success of these algorithms relies on the probability that a random spanning tree can be split in such a way being relatively high. It is therefore of great algorithmic interest to understand the likelihood that a random spanning tree can be split into two equal-sized pieces in real-world dual graphs.  

Furthermore, the probability a random spanning tree can be split into balanced pieces is closely related to the probability a random spanning forest is balanced~\cite{cannon2024sampling}; this is the probability with which a rejection filter applied to the up-down Markov chain accepts~\cite{charikar2022complexity}. This provides additional motivation to understand this quantity in real-world graphs: since the up-down Markov chain produces samples in polynomial time, if the probability a random spanning tree of the graph can be split into $k$ pieces is also polynomial, this would give a polynomial time sampling algorithm for connected balanced partitions in real world settings, which has thus far remained elusive. 

Empirical splittability properties of dual graphs are the main focus of Section~\ref{sec:split}. 

\subsubsection{Results on Grid Graphs}

%In searching to provide theoretical underpinnings for sampling algorithms, researchers have turned to simplified settings.  Most frequently, this has meant looking at an algorithm's properties when the underying dual graph is a grid, which is the natural first simplification to consider. 

In attempting to provide rigorous results regarding sampling algorithms, researchers have largely turned to studying how the algorithms behave on grid graphs. 

One of the most basic question to ask about a Markov chain sampling algorithm is irreducibility: do the moves of the chain suffice to reach every possible districting plan? For recombination Markov chains, the answer is known to be yes when the dual graph is a triangular subgraph of the triangular lattice~\cite{irredj} or a subgraph of the square grid~\cite{akitaya2026redistricting}.  However, questions related to the irreducibility of recombination remain unknown in nearly all other settings. 

Another natural question to consider is the mixing time of a Markov chain: how many steps are needed before the samples produced are sufficiently random?  It's known there exists a grid subgraph on which recombination Markov chains require exponential time to mix~\cite{charikar2022complexity} and a (different) grid subgraph on which Glauber dynamics -- a particular instantiation of a flip Markov chain -- also requires exponential time to mix.  Mixing of both flip and ReCom Markov chains on real-world dual graphs remains an open question, however.  

The up-down Markov chain is known to mix in polynomial time, producing a sample in time $O(n \log^2 n)$ for all dual graphs with $n$ vertices. However, the key question here is whether the partitions produced have equally-sized districts.  Recently it was shown that a polynomial fraction of $k$-partitions are balanced (for constant $k$) in both grids and sufficiently large grid-like lattices~\cite{cannon2024sampling}. Here the characterization of grid-like graphs was chosen so as to enable the theoretical results to hold, not to mimic the properties encountered in real-world dual graphs. 

While grid graphs make a logical starting point for exploration of the properties of various sampling algorithms, they are not a broad enough graph class to encompass the varied structures seen in real-world dual graphs.  Part of our work considers how similar dual graphs really are to square/triangular lattices, which can provide insights into how relevant these theoretical results are to real-world applications of the various sampling algorithms.  

Another main goal of our work is to provide definitions of graphs classes -- and random graph models -- that more closely resemble dual graphs.  The hope is that by using the graph classes, researchers will be able to provide theoretical guarantees about sampling algorithms in settings that more closely resemble how these algorithms are used in practice.

\subsection{Overview}

Section~\ref{sec:props} considers the properties of census dual graphs, mainly at the census tract and census block group levels. This includes basic properties related to degree, connectivity, and planarity, as well as more nuanced properties such as their number of spanning trees and the likelihood a uniformly random spanning tree can be split into equal-sized parts. 

Section~\ref{sec:models} presents and evaluates several random graph models.  The first class of models randomly perturbs grid graphs, and interpolates between the square and triangular grids.  The remaining models connect Poisson point clouds in various ways, including by adding and then perturbing the edges of the Delauney triangulation. 

Section~\ref{sec:conclusion} outlines next steps for further analysis.

\section{Properties of Census Dual Graphs}

\label{sec:props}

\subsection{The Data}\label{sec:data}

For this project, we analyzed dual graphs corresponding to various geographic units across multiple states. The data was provided by Daryl DeFord who created them based on shapefiles from Redistricting Data Hub (\url{https://redistrictingdatahub.org/}) with certain states excluded due to data formatting issues. The datasets provided were:

\begin{itemize}
    \item Counties: All states except Nevada (49 states total)
    \item Census tracts: All states except Nebraska, Nevada, and Wisconsin (47 states total)
    \item Census block groups: All states except Nebraska, Nevada, and Wisconsin (47 states total)
    \item Census blocks: All states except Nevada (49 states total)
\end{itemize}

\noindent Our analysis largely focuses on the tract- and block group-level data, as county graphs were deemed too small to provide interesting insights, and the block-levels graphs were so large (an average of almost 165,000 vertices, with a maximum of over 500,000 vertices) as to create computational bottlenecks.

\subsection{Degree}

As introduced earlier, data was available at four geographic levels: counties, census tracts, census block groups, and census blocks. The preliminary analysis for creating random graph models on dual graphs was done at the census tract and block group levels. County graphs were excluded because counties are too large and few to be interesting, while blocks graphs were deemed too computationally intensive to analyze comprehensively. 

For the selected tract and block group levels, the average, median, and maximum vertex degrees were computed across all available U.S. state-level graphs as seen in Table~\ref{tab:random}. These empirical values will also later serve as reference points for model evaluation. %Ideally, generated graphs should approximate an average degree of 5.4, a median degree of 5, and a maximum degree in the mid 20s.

\begin{table}
    \centering
    \begin{tabular}{|c|c|c|c|c|c|}
        \hline
        \textbf{Map Type} &  \makecell{\textbf{Average} \\ \textbf{Degree}} & 
        \makecell{\textbf{Median} \\ \textbf{Degree}} & 
        \makecell{\textbf{Max} \\ \textbf{Degree}} & \textbf{Connected} & 
        \textbf{Planar} \\ \hline
        Tract & 5.398 & 5.000 & 21.000 & 0.894 & 0.745 \\
        Block group & 5.441 & 5.000 & 27.574 & 0.894 & 0.702 \\
        \hline
    \end{tabular}
    \caption{Analysis of properties of census tract and block group dual graphs across the U.S.. The degree quantities were computed for each available graph at the census block group and tract levels, and then averaged over all graphs at the same level of geography.  The Connected and Planar columns indicate what fraction of graphs at each level of geography had those properties. Answers are rounded to three decimal places. }
    \label{tab:random}
\end{table}

\subsection{Connectivity and Planarity}

Additionally, we analyzed the connectivity and planarity of the dual graphs as seen in Table~\ref{tab:random}. Dual graphs of any contiguous land area should be connected. However, islands or other geographical attributes may cause graphs to be disconnected.  This can vary across different levels of geography; a state's county graph may be connected, but it's tract graph might not, for example. We found that approximately 90\% of census tract and census block group graphs are connected (specifically, 42 our of our 47 graphs at each level). While the connectivity is the same at both the census tract and block group levels for all our data, this is not guaranteed at all levels. For example, states like California and Florida have connected county-level graphs, but their higher-resolution counterparts (tracts and block groups) contain disconnected components.

However, with the exception of Hawaii, a state's dual graph at the tract and block group levels nearly always consists of one large component, plus additional components with five or fewer vertices.  Because of this, for some of our analysis that requires connected graphs (for example, when studying splittability of spanning trees in Sec.~\ref{sec:split}), we instead focus only on the largest connected component. See Table~\ref{tab:comp_sizes}.

\begin{table}

\begin{centering}

\begin{tabular}{|l|l|l|l|}
\hline
State & County & Tract & Block Group \\\hline
 CA & conn & [9126, 2, 1] & [25600, 5, 1, 1] \\\hline
 FL & conn & [5159, 1] & [13387, 1] \\\hline
 HI & [2,1,1,1] & [329, 60, 43, 18, 5, 3, 2, 1] & [773, 147, 96, 49, 10, 3, 2, 2, 1]\\\hline
 NY & conn & [5410, 1] & [16069, 1]  \\\hline
 RI & conn & [248, 2] & [789, 3] \\\hline
    \end{tabular}

\end{centering}
    
    \caption{For all disconnected dual graphs at the Tract and Block Group level, the sizes of their connected components at each level of geography. A designation of conn means the graph is connected. }
        \label{tab:comp_sizes}

\end{table}

Dual graphs of geographic maps are often planar, since they are derived from non-overlapping spatial regions. However, in practice, certain geographic features, such as adjacencies across bodies of water or boundary points where multiple regions meet, can result in non-planar graphs.  We also found that nearly three-quarters of census tract and census block group dual graphs are planar; see Table~\ref{tab:random}.

\subsection{Connectivity via spanning tree count}

\begin{figure}
\begin{center}
\includegraphics[scale=0.75]{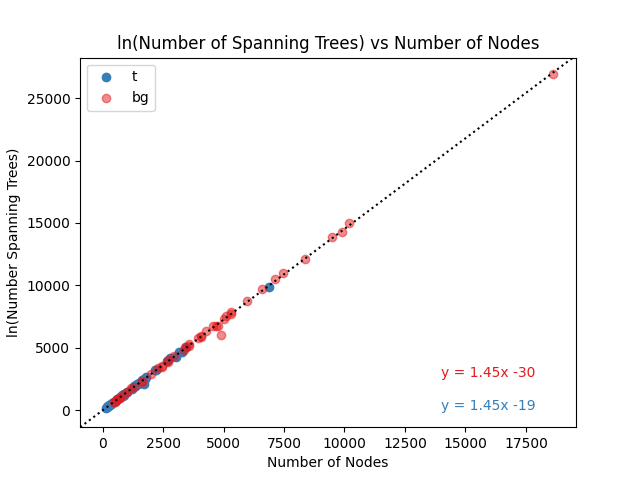}
\caption{A plot showing $\ln(N_{ST}(G))$ vs. $|V(G)|$ for all connected census tract and block group graphs}
\label{fig:t_bg_nodes_lnst}
\end{center}
\end{figure}

As graphs with a larger number of spanning trees tend to be more well-connected, we study the number of spanning trees of dual graphs with the aim of capturing essential aspects of their structure. If $N_{ST}(G)$ is the number of spanning trees of graph $G$, we begin by considering $\ln(N_{ST}(G))$.  These values for all track and block group graphs that are connected (that is, that have at least one spanning tree) are plotted in Figure~\ref{fig:t_bg_nodes_lnst}.

We see in this figure that there is a clear, strong linear relationship between a dual graph's number of vertices and the log of its number of spanning trees.  We therefore choose to focus instead on the quantity $\ln(N_{ST}(G))/|V(G)|$, which we call the {\it spanning tree constant}. The spanning tree constant captures the rate at which the number of spanning trees grows as the graph becomes larger. This quantity approaches a fixed value for certain graphs, such as grid graphs, as the number of vertices increases \cite{tappSTBounds}.

%\todo{rewrite this paragraph}
%To compare graphs of different sizes, researchers often use the spanning tree score, defined as the natural logarithm of the number of spanning trees, with higher scores corresponding to more compact plans \cite{duchin2023discrete}. The spanning tree score is calculated for each district, and the total score for a districting plan is found by summing the scores of all individual districts. In some cases, the score is normalized by dividing by the number of vertices in the graph \cite{cannon2022spanningtreemethodssampling}, which we refer to in this paper as the spanning tree constant. 

We analyzed the spanning tree constant of census tracts and census block groups. The spanning tree asymptote for dual graphs (if it exists) appears to be around 1.45, which is also the slope of the line-of-best-fit for $\ln(N_{ST})$ vs. $|V(G)|$. The average spanning tree constant is 1.43. 
For context, this asymptotic value is about 1.166 for square lattices and 1.615 for triangular lattices~\cite{Wu_1977}, so our experimental results fall in between the two (for more on this, see Section~\ref{sec:grids}). 
%These values suggest a moderate level of compactness and connectivity. 
Both the asymptote and the average spanning tree constant are important to consider when evaluating the models presented in Section~\ref{sec:models}.  The spanning tree asymptote illustrates the behavior of the graphs as the number of vertices increases, while the average spanning tree constant captures the typical compactness of graphs at their observed sizes.

\begin{figure}
\begin{center}
\includegraphics[scale=0.75]{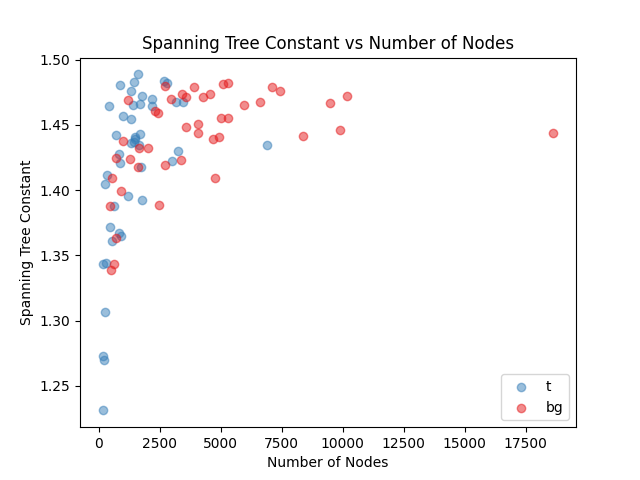}
\caption{Spanning tree constant for census tract and block group dual graphs}
\label{fig:st_cons_real_data}
\end{center}
\end{figure}

\FloatBarrier

\subsection{Splittability}
\label{sec:split}

%\subsubsection{Motivation}
%While existing methods for redistricting analysis provide a robust computational framework for detecting gerrymandering, they suffer from poor scalbility and lack polynomial time guarantees. As a result, generating large and diverse enough ensembles becomes comutaitonally expensive, restircitng the applicability of enselmble-based analysis for for large-scale dual graphs, such as those representing census blocks and tracts.

%In this work, we investigate an alternative graph-based approach grounded in the uniform distribution over spanning trees. Specifically, we examine whether uniformly random spanning trees can be split into components with desired properties. Our empirical results demonstrate that such trees can, with high probability, be split in a fraction of polynomial time, suggesting a more scalable and computationally tractable framework for redistricting analysis. 

%\subsubsection{Approach}
Studying the likelihood a random spanning tree can be split into balanced pieces is key to the polynomial-time sampling guarantee of the algorithm proposed in~\cite{charikar2022complexity} and analyzed for grids and grid-like graphs in~\cite{cannon2024sampling}: If a random spanning tree of a graph with $n$ vertices has at least a $1/poly(n)$ probability that it can be split into balanced pieces, then a valid sample is produced in polynomial time. Thus understanding the splittability properties of dual graphs is of central importance in understanding the efficiency of sampling algorithms for connected balanced partitions of dual graphs. 

\begin{figure}
\begin{center}
\includegraphics[scale=0.4, trim = 0 20 0 39, clip]{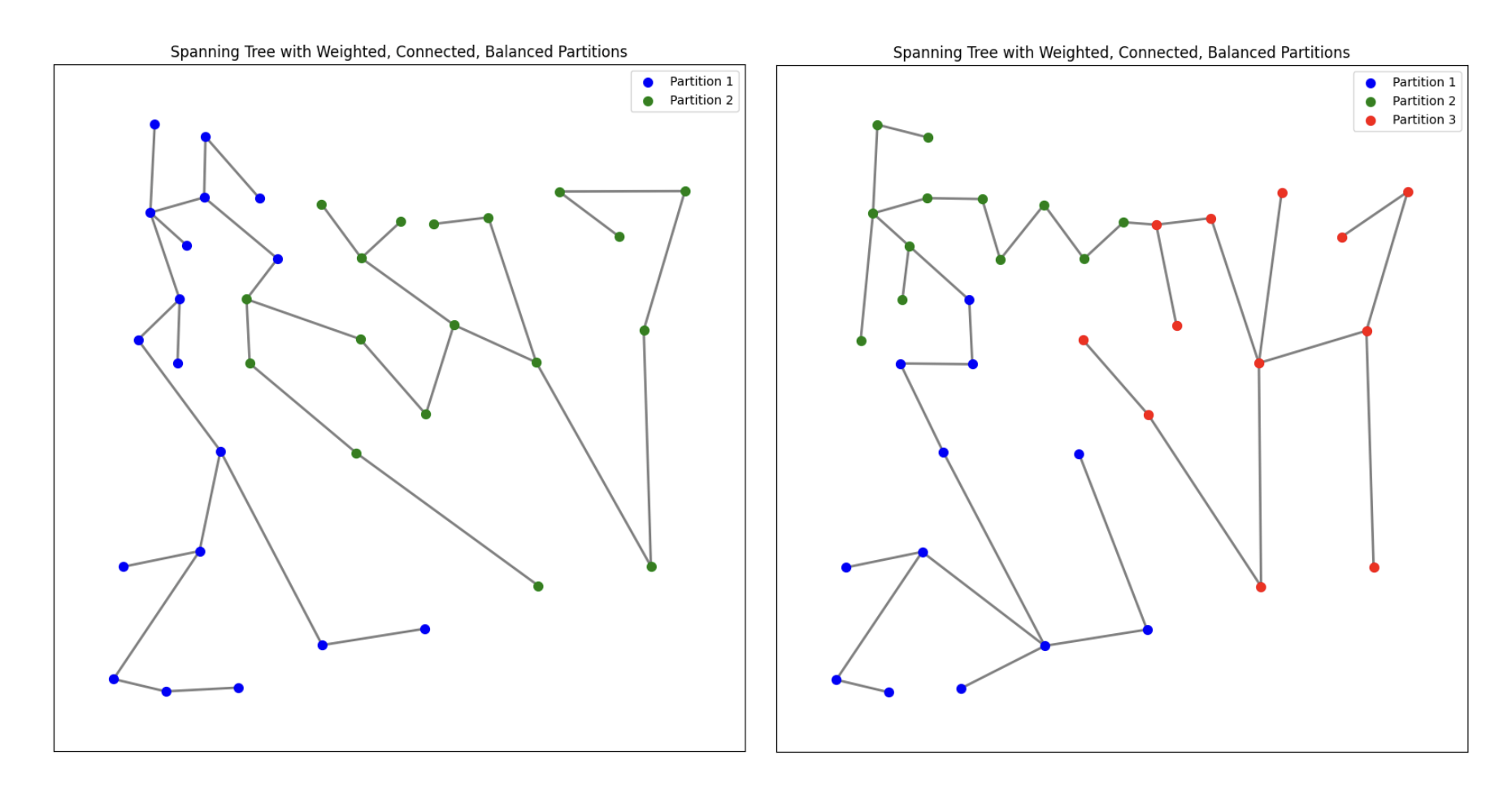}
\caption{For the Oregon counties dual graph, a spanning tree that can be split into two balanced pieces (left) and a spanning tree that can be split into three balanced pieces (eight)}
\label{fig:partitions}
\end{center}
\end{figure}

Though in real-world graphs it is splittability with respect to {\it population} that is most relevant, this initial exploration focuses on the easier question of splittability with respect to {\it vertices}; said another way, we make the simplifying assumption that all dual graph vertices have the same population. Computational restrictions also led us to focus mainly on partitions into $2$ or $3$ balanced pieces, and to include county data in our analysis.  We will use $k$ to denote the number of pieces in the partitions we consider, that is, the number of balanced pieces we are trying to split the random spanning trees into. 
See Figure~\ref{fig:partitions} for examples of balanced connected partitions on the Oregon counties dual graph for $k = 2$ and $k = 3$.

% We're not going to get into this much detail about the methods. 
%The first key idea is to study the probability of splitting a uniformly random spanning tree into $2$-balanced, connected subsets using Wilson's algorithm and Breadth-First Search. For an arbitrary root vertex, Wilson's algorithm generates a uniformly random spanning tree on a connected, undirected graph using a series of loop-erased random walks from arbitrary starting points to the component containing the root \cite{cannon2024sampling}. In this context, {\it random} means that each step in the walk selects a neighboring vertex with equal probability. 

%Once a uniform random spanning tree has been generated, we apply Breadth-First Search (BFS), a standard graph traversal algorithm, to attempt a balanced split of the tree. Together, Wilson's algorithm and BFS provide a framework for sampling $k$-connected, balanced spanning trees from an input dual graph, assuming such a partition exists. See Figure~\ref{fig:partitions} for examples of 2- and 3-balanced connected partitions on the Oregon counties (cnty) dual graph.

Computationally, for all dual graphs at the county, tract, and block group levels, we generated uniformly random spanning trees using Wilson's algorithm~\cite{wilson1996graph} and checked whether they were splittable into the desired number of pieces using a variant of breadth-first search. For dual graphs that are not connected, we performed this process on the graph's largest connected component. 

To quantify the probability of obtaining a splittable uniformly random spanning tree (UST), we employ the Gamma Bernoulli Approximation Scheme (GBAS)~\cite{Huber2021}. This offers a probabilistic framework for estimating the number of successful trials needed to ensure, with 95\% confidence, that our approximation is statistically reliable. This approach estimates the parameter $p$ of a Bernoulli-distributed random variable by constructing an unbiased estimate $\hat{p}$ such that the relative error $(\hat{p}/p-1)$ follows a known distribution independent of $p$. This property enables accurate estimation of the probability of generating a splittable UST under repeated sampling, without relying on limiting approximations. In particular, we aim for estimates that, with $95\%$ confidence, are with 25\% of the true value; as the GBAS prescribes, this means we must test random spanning trees for splittablity until we have seen 178 successes, where a success occurs when the randomly spanning tree under consideration can be split. These values were chosen to balance accuracy with computational feasibility. 
 
\subsubsection{Splittability Estimates}

We conducted repeated trials on U.S. state dual graphs at the county, tract, and block group level. At the county level, we excluded Hawaii and Delaware, which contain only 5 and 3 counties, respectively, and whose county dual graphs are always splittable. We continued each experiment until 178 splittable random spanning trees were found. Then, 178 divided by the total number of attempts needed to see these successes produces an unbiased estimate $p$ for each dual graph of the probability that a random spanning tree of the dual graph is splittable. We were particularly interested in how this probability relates to the number of vertices in the graph. 

\begin{figure}
\begin{center}
\includegraphics[scale=0.6]{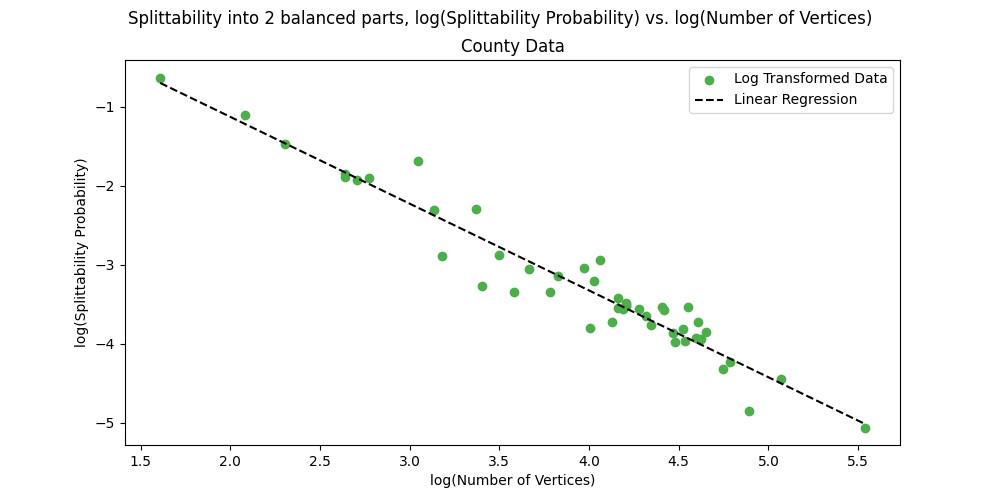}
\caption{Logarithmic transformation of county data, with regression line  {$\log(p)  = -1.0991 \log(n) + 1.0702$}, for $p$ the $2$-splittability probability estimate and $n$ the number of vertices 
\label{fig:cnty} }
\end{center}
\end{figure}

\begin{figure}
\begin{center}
\includegraphics[scale=0.6]{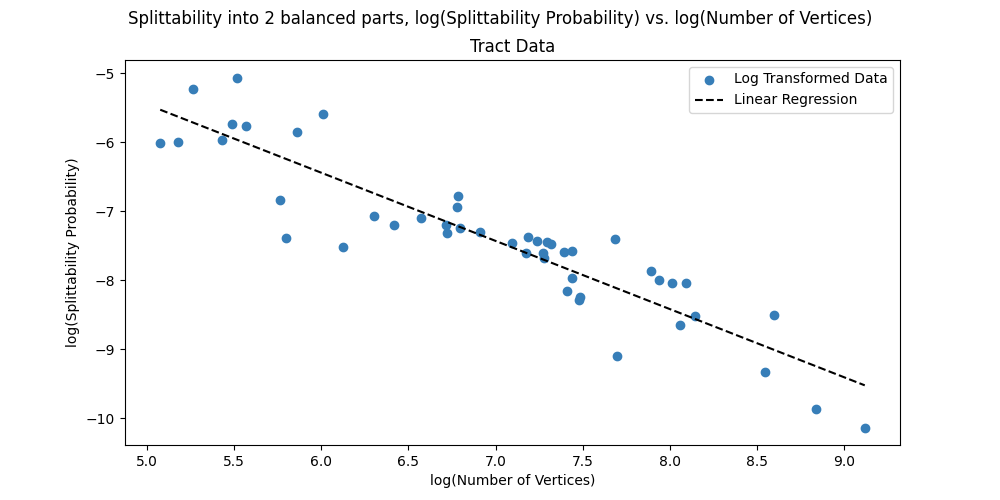}
\caption{Logarithmic transformation of tract data, with regression line  {$\log(p) = -0.9887\log(n) - 0.5089$}, for $p$ the $2$-splittability probability estimate and $n$ the number of vertices
\label{fig:t_split} }
\end{center}
\end{figure}

\begin{figure}
\begin{center}
\includegraphics[scale=0.6]{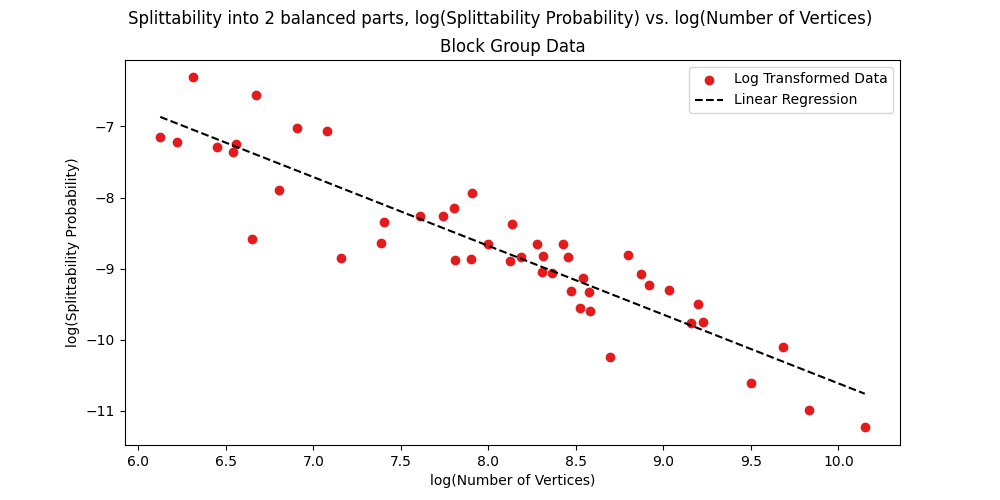}
\caption{Logarithmic transformation of block group data, with regression line {$\log(p) =  -0.9669 \log(n) - 0.9440$}, for $p$ the $2$-splittability probability estimate and $n$ the number of vertices. 
\label{fig:bg_split} }
\end{center}
\end{figure}

\underline{2-splittability}: Figure~\ref{fig:cnty} shows the relationship, for the county dual graphs, of logarithm of the number of vertices $n$ and the logarithm of the estimated 2-splittability probability $p$. The same quantities for tract and block group data are shown in Figures~\ref{fig:t_split} and~\ref{fig:bg_split}, respectively.

Notably, in all three cases there appears to be a linear relationship between these two log-transformed quantities. The resulting regression equations gives
\begin{align*}
    &\log(p) = -1.0991 \log(n) + 1.0702 \text{ for counties} \\
    &\log(p) = -0.9887\log(n) - 0.5089 \text{ for tracts} \\
    &\log(p) =  -0.9669 \log(n) - 0.9440 \text{ for block groups}
\end{align*}
These models have $R^2$ values of $0.9337$, $0.8289$, and $0.8260$, respectively, indicating a very strong linear relationship between these two logarithmic quantities. The strong linear fit on the log-log scale suggests the raw data is well approximated by an inverse polynomial model.
Exponentiating both sides, we obtain the relationships
\begin{align*}
    &p = 2.9160 n^{-1.0991} \text{ for counties} \\
    &p = 0.6011 n^{-0.9887} \text{ for tracts} \\
    &p = 0.3891 n^{-0.9669} \text{ for block groups}
\end{align*}
As we see here, the estimated exponential relationship for the original data has exponents of approximately $-1.0991$, $-0.9887$, and $-0.9669$. Notably, an exponent of exactly $-1$ would correspond to a perfect inverse linear relationship, with the splittability probability scaling precisely as $\frac{1}{n}$ for graphs with $n$ vertices. Given that all estimates are close to $-1$, we infer the raw data closely approximates, or at least trends toward, an inverse linear model. In~\cite{cannon2024sampling}, authors proved a lower bound on the order of $1/n^2$ for splittability probabilities in grid graphs; our results suggest $1/n$ is likely much closer to the truth.

%Despite initially performing this analysis separately on counties, tracts, and block groups, we also see that despite the variance in model coefficients, the data for all three geographies, when combined, also follows an overall linear trend
%(Figure~\ref{fig:all_2_split}). The resulting regression line for the combined geographies is $\log(p) =  - 1.2622 \log(n) + 1.5123 $, with a $R^2$ value of $0.9637$. 
%\begin{figure}
%\begin{center}
%\includegraphics[scale=0.5]{imgs/all_2_log_v_log.png}
%\caption{Logarithmic transformation of county, tract, and block group data, with overall regression line  {$\log(p) =  - 1.2622 \log(n) + 1.5123 $}, for $p$ the $2$-splittability probability estimate and $n$ the number of vertices. 
%\label{fig:all_2_split} }
%\end{center}
%\end{figure}

\begin{figure}
\begin{center}
\includegraphics[scale=0.6]{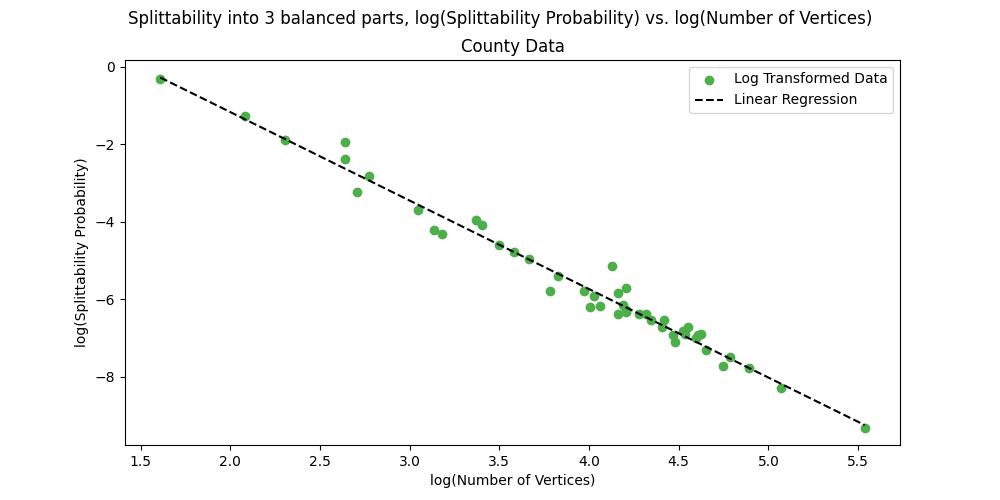}
\caption{Logarithmic transformation of county data, with overall regression line  {$\log(p) = -2.2818 * \log(n) + 3.3940 $}, for $p$ the $3$-splittability probability estimate and $n$ the number of vertices 
 }
\label{fig:cnty3}
\end{center}
\end{figure}

\begin{figure}
\begin{center}
\includegraphics[scale=0.6]{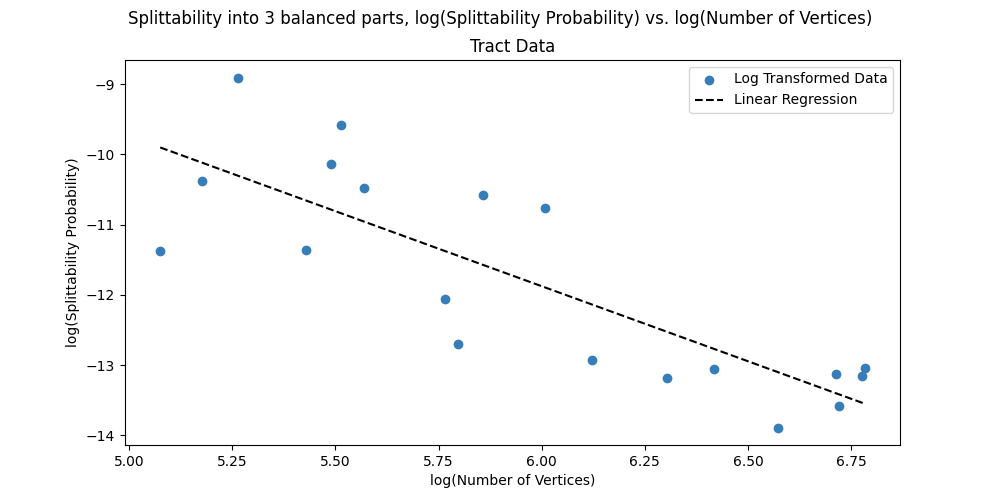}
\caption{Logarithmic transformation of tract data, with overall regression line  {$\log(p) = -2.1363 * \log(n) + 0.9416 $}, for $p$ the $3$-splittability probability estimate and $n$ the number of vertices 
 }
\label{fig:t3}
\end{center}
\end{figure}

\vspace{3mm}
\underline{3-splittability}:  We also moved beyond 2-splittability, as computational resources allowed. 
We estimated 3-splittability of county graphs for all states, and 3-splittabliity of tract graphs for all states for which 178 successes could be found in less than one week on a cluster computer, which turned out to be 19 states.  The resulting figures on a log-log scale for counties and tracts are shown in Figures~\ref{fig:cnty3} and~\ref{fig:t3}, respectively. We performed linear regression for each geography, and the resulting equations are: 
\begin{align*}
    &\log(p) = -2.2818 \log(n) + 3.3940 ; \hspace{5mm}  p = 29.7859 n^{-2.2818} \hspace{5mm} \text{ for counties, $k = 3$} \\
    &\log(p) = -2.1363\log(n) + 0.9416 ; \hspace{5mm}  p = 2.5640 n^{-2.1363} \hspace{5mm} \text{ for tracts, $k = 3$}
\end{align*}
The $R^2$ values are $0.9785$ and $0.6706$ for counties and tracts, respectively. 

\begin{figure}
\begin{center}
\includegraphics[scale=0.6]{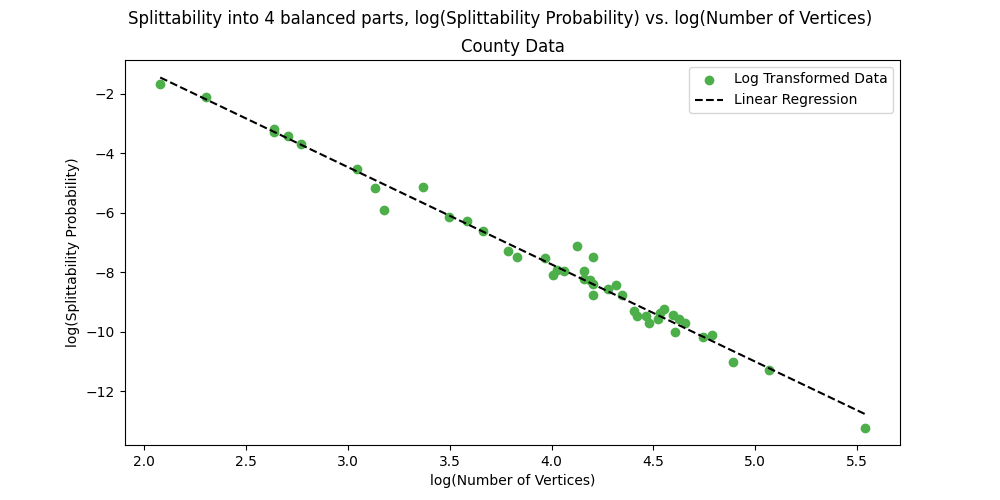}
\caption{Logarithmic transformation of county data, with overall regression line  {$\log(p) =  -3.2739  \log(n) + 5.3569 $}, for $p$ the $4$-splittability probability estimate and $n$ the number of vertices 
 }
\label{fig:cnty4}
\end{center}
\end{figure}

\vspace{3mm}
\underline{4-splittability}: Finally, we estimated 4-splittability probabilities for all county graphs, and  the resulting estimated 4-splittability probabilities on a log-log scale are shown in Figure~\ref{fig:cnty4} (for all states except Alaska, whose county graph has no 4-splittable spanning trees and thus a resulting probability of~0). We performed linear regression, yielding: 
\begin{align*}
    &\log(p) = -3.2739 \log(n) + 5.3569 ; \hspace{5mm}  p = 212.0684 n^{-3.2739} \hspace{5mm} \text{ for counties, $k = 4$} 
\end{align*}
The $R^2$ value for this regression was $0.9829$. 

\vspace{3mm} 
We summarize the estimated exponents of the inverse exponential relationship between the number of vertices and the splittability probability in Table~\ref{tab:split}.  We see that the exponent has limited dependence on the type of geography, but significant dependence on the number of parts $k$ we are looking to split a random spanning tree into.  Based on this data, a reasonable hypothesis might be to predict the exponential dependence $p = \Theta(n^{-(k-1})$, that is, predict $p \sim 1/n$ for $k = 2$, $p \sim 1/n^2$ for $k = 3$, and $p \sim 1/n^3$ for $k = 4$, where we are using $\sim$ to denote approximate rates of growth. Of course, these hypotheses do not perfectly  match our observed values, and there may be additional (lower-order?) terms contributing to these quantities. Additional computational experiments and more accurate estimates of $p$ could help refine these hypotheses. 

\begin{table}
\centering
\begin{tabular}{|c|c|c|c|}\hline
& $k = 2$ & $k = 3$ & $k = 4$ \\\hline
Counties & -1.0991 & -2.2818 & -3.2739 \\\hline
Tracts & -0.9887 & -2.1363 & $-$ \\\hline
Block Groups & -0.9669 & $-$ & $-$ \\\hline
\end{tabular}
    \caption{For all three geographies we considered, the exponent $b$ in our modeled equation $p = a n^b$, where $n$ is the number of vertices and $p$ is the probability a random spanning tree can be split into $2$, $3$, or $4$ parts, respectively. We were unable to compute estimates for the combinations marked $-$ due to the large amount of computation required. }
    \label{tab:split}
\end{table}

\FloatBarrier

\section{Random Graph Models}

\label{sec:models}

In this section, we further investigate the structure of dual graphs by considering several random graph models and seeing how they compare to dual graphs along key statistics. 

\subsection{Perturbing Grids}
\label{sec:grids}

The most basic graph to construct is a grid graph, or square lattice graph. This is a rectangular graph which has vertices at integer points that are connected to neighboring vertices with vertical and horizontal edges. The vertices in the interior of the grid will all have degree 4, and vertices along the outside have degree 3, except for the four corner points which will have degree 2. Therefore the maximum degree of a grid graph is 4, and the average degree approaches 4 with larger grids.

We can also construct a triangular grid graph from the square grid graph by adding the diagonals edges from all vertices $(x,y)$ to their diagonal neighbor $(x+1,y+1)$, with the exception of vertices along the top and far right boundaries which have no such neighboring vertex. In a triangular grid graph, the interior vertices will have degree 6, vertices along the outside will have degree 4, two corner points will have degree 3, and the other two corner points will have degree 2. This makes the maximum degree of the graph 6, and the average degree approaches 6 with larger graphs. 

\begin{figure}
\begin{center}
\includegraphics[scale=0.4]{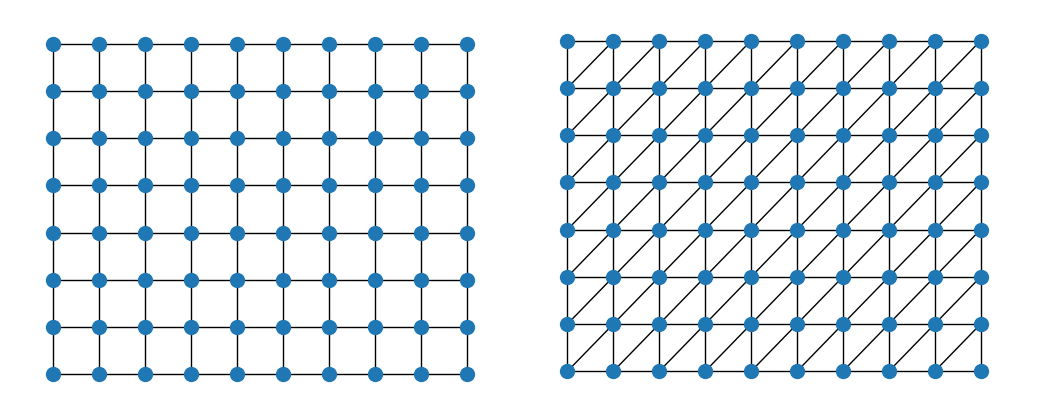}
\caption{Examples of Square and Triangular grid graphs}
\label{fig:sq_tri_grids}
\end{center}
\end{figure}

Examples of both square and triangular grids can be found in Figure~\ref{fig:sq_tri_grids}. Most previous work in the space of redistricting which seeks a simplified setting in which to work has considered either the square or triangular grids~\cite{cannon2024sampling,irredj,tappSTBounds, charikar2022complexity}.

\begin{figure}
\begin{center}
\includegraphics[scale=0.7]{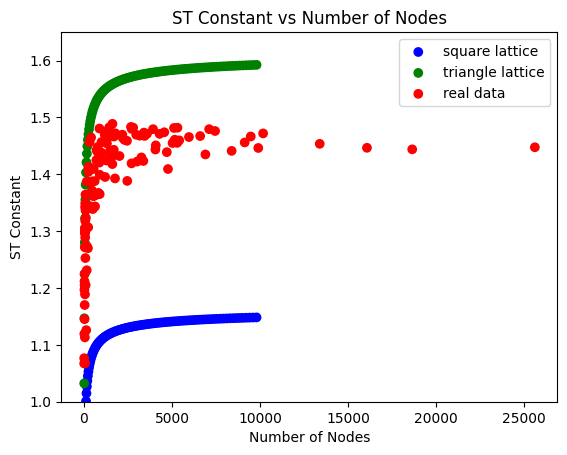}
\caption{Computed spanning tree constants vs. number of vertices for dual graphs (red, middle), square grids (blue, bottom), and triangular grids (green, top)}
\label{fig:sq_tri_vs_real}
\end{center}
\end{figure}

Recall we defined the {\it spanning tree constant} of a graph $G$ to be $\ln(N_{ST}(G))/|V(G)|$, and experimentally found it to approach an asymptote of around 1.45 for our dual graphs. 
We can also calculate the spanning tree constant for square and triangular lattices of various sizes and compare it to real data, which we do in Figure~\ref{fig:sq_tri_vs_real}. We find the spanning tree constant for our real data falls somewhere between that of the square and triangular grid. To find a better model to approximate real data, we can construct a random graph which is something between a square and triangular grid, which we call a randomly perturbed grid. 

\begin{figure}
\begin{center}
\includegraphics[scale=0.4]{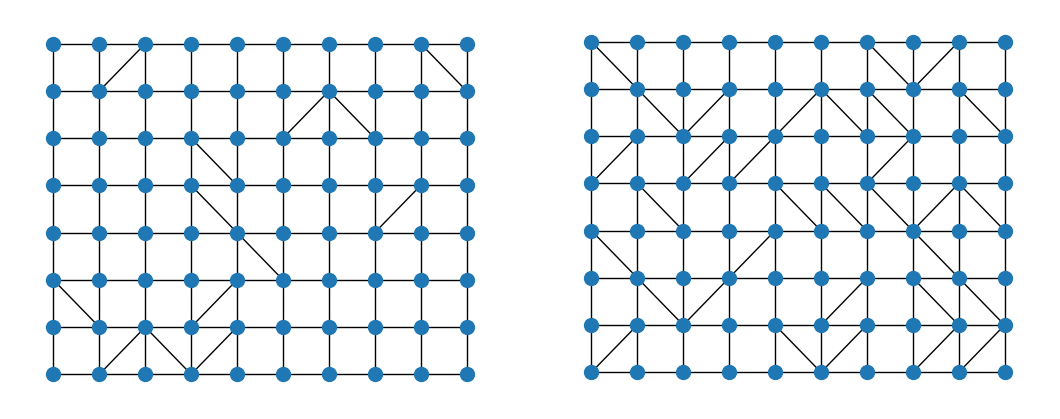}
\caption{Randomly perturbed grids with probabilities $p=0.25$ (left) and $p=0.5$ (right)}
\label{fig:pert_grids}
\end{center}
\end{figure}

To construct a randomly perturbed grid, we use a square grid graph as a base and randomly add diagonal edges. With a certain probability $p$, we can add a diagonal edge to each square in the grid, and the edge has a $50\%$ probability of going in either direction. Note that in the triangular lattice, our diagonals are all going in the same direction. Allowing for diagonals in either direction gives us more variety in the degrees of our vertices, with a maximum degree of up to 8. Two examples of randomly perturbed grids are shown in Figure~\ref{fig:pert_grids}.

\begin{figure}
\begin{center}
\includegraphics[scale=0.7]{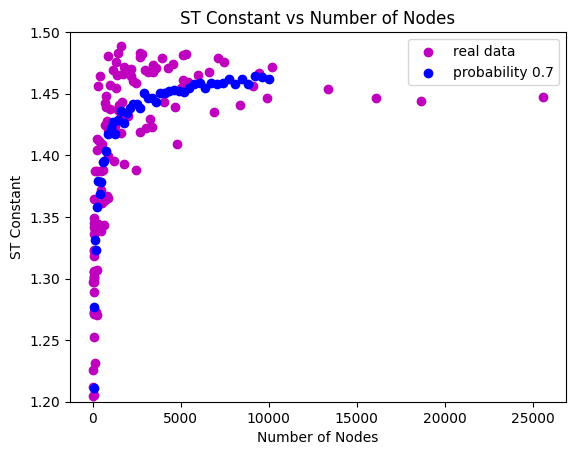}
\caption{Compare spanning tree constant for real data to perturbing grid with probability $p=0.7$.}
\label{fig:pert_grid_vs_real}
\end{center}
\end{figure}

We experimented with different probabilities $p$ of adding a random diagonal to a cell of a square grid.  We found that the spanning tree constant for perturbing grids is closest to our real data when we use probability $p=0.7$; see Figure~\ref{fig:pert_grid_vs_real}. This suggests this model of randomly perturbed grids most closely resembles our dual graphs, and researchers looking for simplified settings to study who want to move slightly beyond grid graphs should consider this model.

\subsection{Constructing Random Graphs}\label{sec:RGs}

We now move beyond grid-based random graph models, to models that generate random point clouds in the plane and then connect those points following various sets of guidelines. 

Formally, to construct each random graph model, we begin by generating a set of vertices randomly embedded in a 2D space. Given a specified $n$ number of vertices and a fixed random seed, each vertex is assigned random coordinates within the unit square $[0,1]\times[0,1]$. Once all vertex coordinates are established, edges are assigned based on the specific criteria of each model. 

In our model exploration, we began with a model that adds edges probabilistically, with a higher likelihood of connecting nearby vertices and a lower likelihood for distant pairs, favoring local connectivity. Next, we explored a deterministic approach that adds the shortest edges to the graph in order of increasing length. Finally, we investigated Delaunay triangulations (see Section~\ref{sec:background_dt}) and analyzed how different strategies for edge removal and addition influence both the average degree and the spanning tree constant, both its asymptotic behavior and its average value. 
To see a summary of all twelve models we considered, see Table~\ref{tab:model_summary_des}. 

\begin{table}
    \centering
    \begin{tabular}{c|c|p{8cm}}
        \hline
        & \textbf{Model} & \textbf{Description} \\ \hline
        \multirow{3}{*}{\makecell{Connecting \\ Random Point \\ Clouds}} & 1  & Base, distance probability \\
        & 2 & Add $\frac{5.4}{2}n$ shortest edges \\
        & 11 & Add $\frac{s}{2}n$ shortest edges and remove random edges \\ \hline
        \multirow{9}{*}{\makecell{Delaunay \\Triangulations\\of Random\\Point Clouds}} & 3 & Basic triangulation \\
        & 4 & Triangulation with random removal \\
        & 5 & Triangulation with random removal and random addition \\
        & 6 & Triangulation with adding $n$ shortest edges \\
        & 7 & Triangulation with adding $n$ shortest edges and removing longest edges \\
        & 8 & Triangulation with random removal and adding edges via preferential attachment \\
        & 9 & Triangulation with random removal and adding $n$ shortest edges \\
        & 10 & Triangulation with adding $n$ shortest edges and random removal \\
        & 12 & Triangulation with adding edges via preferential attachment and random removal \\
        \hline
    \end{tabular}
    \caption{Summary of all random graph models we considered.}
\label{tab:model_summary_des}
\end{table}

Our goal is to develop random graph models that replicate key structural properties of real dual graphs derived from U.S. districting data. We evaluate each model based on how closely they match real data regarding average degree, average spanning tree constant, and the asymptotic behavior of the spanning tree constant.
The goal was to create a planar and connected model that reproduced the characteristics of the real data, summarized in Table~\ref{tab:real-data}. After presenting all of our models, we summarize the planarity, connectivity, average degree, median degree, maximum degree, and average spanning tree constant of each model in Table~\ref{tab:results}.

In the following sections, we will discuss the models in various levels of depth, focusing on those that more closely mimicked the desired properties of dual graphs. For additional information and careful consideration of all models, see \cite{af-thesis}. 

\begin{table}
    \centering
    \begin{tabular}{|c|c|c|c|c|c|c|}
        \hline
        \makecell{\textbf{Avg.} \\ \textbf{Degree}} & 
        \makecell{\textbf{Median} \\ \textbf{Degree}} & 
        \makecell{\textbf{Max} \\ \textbf{Degree}} & 
        \makecell{\textbf{Avg. ST} \\ \textbf{Constant}} & 
        \makecell{\textbf{ST} \\ \textbf{Asymptote}} 
        \\ \hline
        5.42 & 5.00 & 24.28 & 1.43 & $\approx$ 1.45 \\
        \hline
    \end{tabular}
    \caption{Characteristics of real-world dual graphs; our target values for random graph models.}
    \label{tab:real-data}
\end{table}

\subsection{Connecting Random Point Clouds: Models 1, 2, 11}
\label{sec:random-pt}

We first present our models that do not use Delauney triangulations. 

\subsubsection{Model One}\label{sec:model1}

The first model explored adding edges based on a given probability where vertices separated by less distance had a higher probability of being connected. The probability of connecting two vertices in the model is defined as $b^{-d}$, where $b$ is a tunable base value and $d$ is the Euclidean distance between the vertices.

For each instantiation of this model, parameter $b$ was chosen to produce an average degree of approximately 5.4, consistent with empirical observations from real-world dual graphs. Because this value depends on both the number of vertices and the random seed, $b$ must be adjusted manually for each configuration. Across different seeds with the same number of vertices, the required base varied.

To reduce the manual tuning of $b$, various transformations were applied to explore potential relationships between $b$ and the number of vertices. See \cite{af-thesis} for further explanation. Since no consistent relationship was identified, subsequent models were pursued for greater efficiency and reliability.

\subsubsection{Model Two}\label{sec:model2}

Model Two constructs graphs by adding the $\frac{5.4}{2}n$ shortest edges, where $n$ is the number of vertices in the graph. This ensures that the resulting graphs achieve the desired average degree of 5.4. The model was then modified to select the largest connected component. While connected, the graphs were never planar across all of our trials. These largest components had an average degree of 5.6, a median degree of 5.41, a maximum degree of 11.18, and an average spanning tree constant of 1.19.

\subsubsection{Model Eleven}\label{sec:model11}

In comparing the average degree and spanning tree constant of each model to those of the real data, Model Two proved to be a close approximation, although it slightly overestimated the average degree and underestimated the spanning tree constant. Model Eleven refines this approach by incorporating two modifications: (1) increasing the number of shortest  edges added and then, (2) randomly removing edges. Rather than adding the $\frac{5.4}{2}n$ shortest edges, this model adds the $\frac{s}{2}n$ shortest edges, and then randomly removes each edge with some probability $p$ chosen to produce the desired average degree. Three combinations of scaling factors and removal probabilities were evaluated; see Table~\ref{tab:model11}. Across all of our trials, the graphs generated were always connected but never planar. 

\begin{table}
    \centering
    \begin{tabular}{|c|c|c|c|c|c|c|c|}
        \hline
        \makecell{\textbf{Model}} & 
        \makecell{\textbf{Scaling} \\ \textbf{Factor $s$}} & 
        \makecell{\textbf{Removal} \\ \textbf{Probability $p$}} & 
        \makecell{\textbf{Average} \\ \textbf{Degree}} & 
        \makecell{\textbf{Median} \\ \textbf{Degree}} & 
        \makecell{\textbf{Max} \\ \textbf{Degree}} &
        \makecell{\textbf{Avg. ST} \\ \textbf{Constant}} &
        \makecell{\textbf{ST} \\ \textbf{Asymptote}} \\ \hline
        11 & 6.8 & 0.2 & 5.48 & 5.29 & 11.37 & 1.27 & $\approx$ 1.3\\
        11b & 9 & 0.4 & 5.43 & 5.23 & 11.59 & 1.30 & $\approx$ 1.35\\
        11c & 14 & 0.6 & 5.46 & 5.24 & 11.9 & 1.36 & $\approx$ 1.45\\
        \hline
    \end{tabular}
    \caption{Summary of the three Model 11 variations we considered}
    \label{tab:model11}
\end{table}

\begin{figure}
\begin{center}
\includegraphics[scale=0.8]{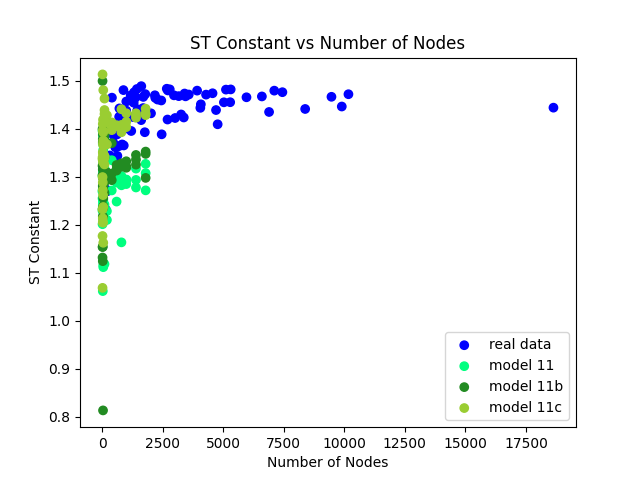}
\caption{A comparison of the spanning tree constant vs. number of nodes for real data (blue) and random graphs generated using Model 11 (greens)}
\label{fig:st_cons_real_data_m11}
\end{center}
\end{figure}

\begin{figure}
\begin{center}
\includegraphics[scale=0.5, trim = 0 60 0 60, clip]{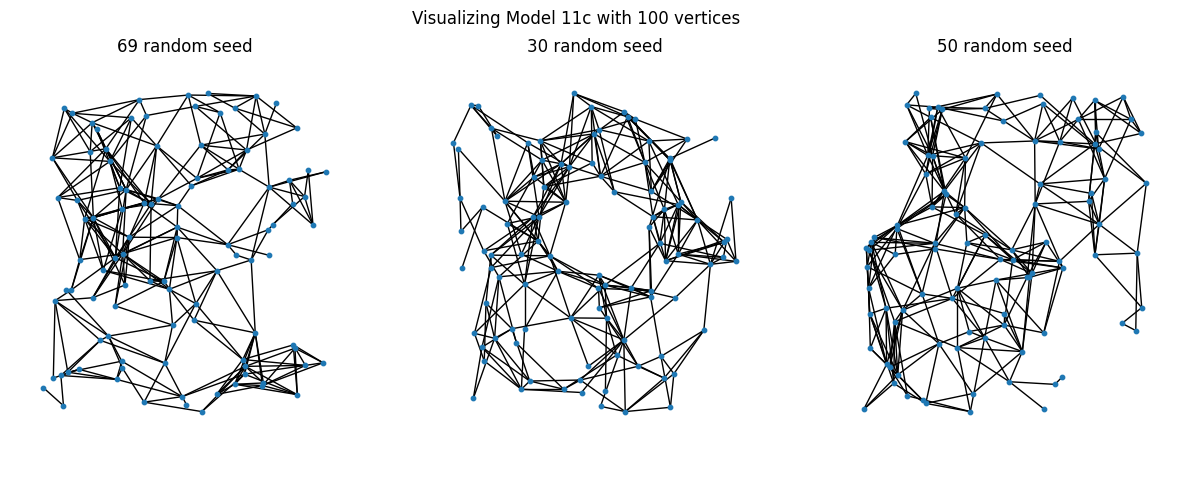}
\caption{Visualizing three different random instances of Model 11c with 100 vertices}
\label{fig:visualize_model11c}
\end{center}
\end{figure}

Model Eleven accurately reproduces both the average degree and the spanning tree constant observed in the real data, with Model 11c performing best.
As seen in Figure~\ref{fig:st_cons_real_data_m11}, though the average spanning tree constant is slightly smaller than real data, the potential spanning tree asymptote for Model 11c most closely resembles that of our real data.  We visualized three instances of Model 11c with 100 vertices in Figure~\ref{fig:visualize_model11c}; the model tends to produce graphs that are densely connected in some regions while more sparsely connected in others, reflecting a realistic balance of local clustering and global dispersion.
Overall, these results suggest that a combination of carefully tuned edge addition and probabilistic edge removal can generate graph structures that closely mirror empirical connectivity patterns. Future work can more carefully examine the precise values of $s$ and $p$ that produce the best fit to the real world data.

\subsection{Background: The Delaunay Triangulation}\label{sec:background_dt}

\begin{figure}
\begin{center}
\includegraphics[scale=0.6, trim = 0 55 0 60, clip]{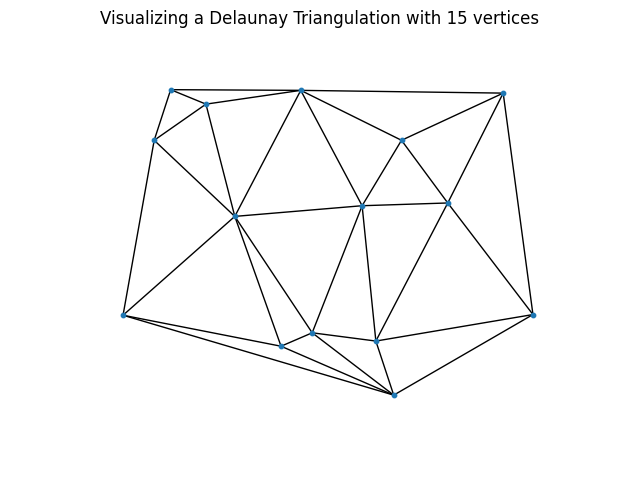}
\caption{Example of a Delaunay triangulation of a point set with 15 points}
\label{fig:visualize_dt}
\end{center}
\end{figure}

Several of the models we consider are based on Delaunay triangulations, a fundamental object widely used in computational geometry. A {\it Delauney triangulation} is a triangulation of a point set such that no point lies inside the circumcircle of any triangle; for a given point set, its Delauney triangulation is unique. An example of the Delauney triangulation of a set of 15 points is given in Figure~\ref{fig:visualize_dt}.

Delaunay triangulations were originally presented in 1934 by Boris Delaunay and more detail can be found in Chapter 2 in {\it Delaunay Mesh Generation} \cite{delaunay1934sphere, delaunayChapter2}. For a given point set, the Delauney triangulation maximizes the minimum angle across all triangles, helping to avoid long, thin triangles and resulting in a mesh that is more compact and regular than many alternative triangulations. In a Delaunay triangulation, each edge $e$ belongs to at most two triangles \cite{delaunayChapter2}. An edge is said to be locally Delaunay if the sum of the angles opposite the edge in the two adjacent triangles is less than or equal to $\pi$. This condition ensures that no vertex of one triangle lies inside the circumcircle of its neighboring triangle.

 %Figure~\ref{fig:dt} demonstrates how flipping edge $e$ from vertices $v_2$ and $v_4$ to vertices $v_1$ and $v_3$ ensures the Delaunay condition. Figure~\ref{fig:visualize_dt} shows a Delaunay triangulation with 15 vertices.

%\begin{figure}
%\begin{center}
%\includegraphics[scale=0.5]{imgs/math_background_egs/delaunay_triangulation.drawio.png}
%\caption{(a) In this quadrilateral, edge \textit{e} is not locally Delaunay, but (b) and (c) illustrate that flipping edge \textit{e} satisfies the Delaunay condition.}
%\label{fig:dt}
%\end{center}
%\end{figure}

One method for constructing a Delaunay triangulation is incremental insertion, which begins with a single triangle and adds points one by one. Following each insertion, the triangulation is updated to maintain the Delaunay condition by flipping edges as necessary.
In our work, we construct Delaunay triangulations of point sets using a different approach. Once given a set of $n$ vertices, their coordinates are converted into a NumPy array and passed to \texttt{scipy.spatial.Delaunay}, a Python implementation of the Qhull algorithm. This computes a Delaunay triangulation by ensuring that no point lies inside the circumcircle of any triangle. The resulting simplices (triangles) are then iterated over to add undirected edges between each pair of vertices within each triangle in order to create a graph structure.

Although Delaunay triangulations are not dual graphs, they share several structural properties such as planarity, connectivity, and compactness. Because of these similarities, they serve as a useful starting point of comparison when analyzing real-world dual graphs based on political or geographical boundaries.

\subsection{Delaunay Triangulations and Perturbations Thereof}
\label{sec:dt}

The remainder of our random graph models begin with the Delauney triangulation of a random point set, and then further perturb this triangulation in a variety of ways. As a starting point, the Delauney triangulation is always a connected and planar graph, though the additional perturbations we make may change this. Recall a summary of all the models we consider is given in Table~\ref{tab:model_summary_des} and a summary of the properties of each model is given in Table~\ref{tab:results}. 

%This geometric structure connects points to form triangles such that no point lies within the circumcircle of any triangle, which results in a planar and connected graph. From the initial structure, subsequent models would add or remove edges as seen fit to achieve the desired properties.

\subsubsection{Model Three}\label{sec:model3}

We first started with a Delaunay triangulation of $n$ points. Unlike Models One and Two, graphs from Model Three are always connected and planar. They feature an average degree of 5.57, a median degree of 5.56, a maximum degree of 9.66, and an average spanning tree constant of 4.29.

\subsubsection{Model Four}\label{sec:model4}

Model Four modifies the Delaunay Triangulation by randomly removing edges, thereby reducing the spanning tree constant while maintaining planarity. Furthermore, by then selecting the largest connected component, the model ensured connectivity. We tested two models, one with a removal probability of 0.2 and another with 0.4. While both brought the spanning tree constant closer to that of the real data, increasing the removal probability decreased the degree metrics too much to be useful.

\subsubsection{Model Five}\label{sec:model5}

Model Five extends Model Four by adding random edges after random removal, aiming to recover lost connectivity while retaining a reduced spanning tree constant. An edge addition probability of 0.05 was used in combination with the same removal probabilities from Model Four. The resulting graphs were always connected but rarely planar. Although we were successful in increasing the average degree, at the number of vertices we considered, the spanning tree constant appears to be diverging. This attribute makes this model unreliable and inaccurate.

\subsubsection{Models Six and Seven}\label{sec:model6&7}

Both Models Six and Seven add the $n$ shortest edges not already in the Delaunay triangulation to enhance local connectivity. Since this will most likely increase the spanning tree constant, Model Seven attempts to counterbalance this by removing the $n$ longest edges, aiming to reduce the constant while preserving local structure. We found that Model Seven was able to produce a spanning tree constant closer to that of real data while maintaining realistic degree distributions, although still too large.

\subsubsection{Model Eight}\label{sec:model8}

Since Model Four produced a spanning tree constant close to the real data but resulted in lower vertex degrees, Model Eight builds on this by adding preferential attachment after randomly removing edges. This technique favors adding edges to vertices with high-degrees. With a removal probability of 0.05 and an edge addition scaling factor of 0.3, this model closely resembled the real data characteristics in terms of degree metrics, as shown in Table~\ref{tab:model8}.

\begin{table}
    \centering
    \begin{tabular}{|c|c|c|c|c|}
        \hline
        \textbf{Average Degree} & \textbf{Median Degree} & \textbf{Max Degree} & \textbf{Avg. ST Constant} & \textbf{ST Asymptote} \\ \hline
        5.49 & 5.4 & 10.16 & 1.45 & $\approx$ 1.6 \\
        \hline
    \end{tabular}
    \caption{Analysis of Model 8}
    \label{tab:model8}
\end{table}

\begin{figure}
\begin{center}
\includegraphics[scale=0.8]{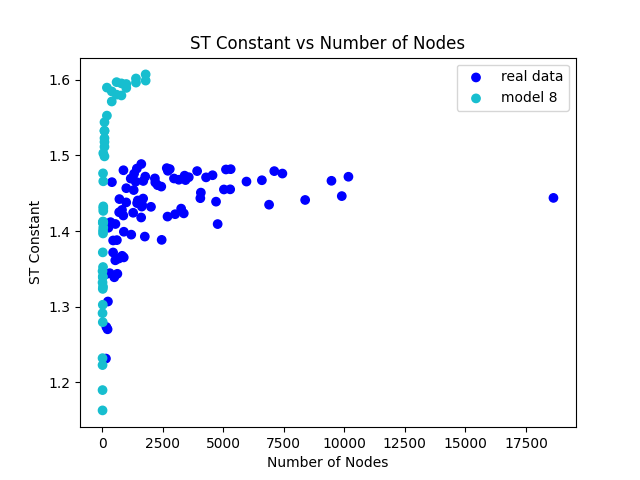}
\caption{Compare spanning tree constant for real data to Model 8 results}
\label{fig:st_cons_real_data_m8}
\end{center}
\end{figure}

While the average spanning tree constant of Model Eight (1.45) is nearly identical to that of the real data (1.43), the asymptote of the spanning tree constant for Model Eight remained too high, as shown in Figure~\ref{fig:st_cons_real_data_m8}. Nevertheless, when considering both the average degree and the average spanning tree constant, Model Eight joins Model Eleven as the two closest overall matches to the real data observed thus far.

\begin{figure}
\begin{center}
\includegraphics[scale=0.5, trim = 0 60 0 60, clip]{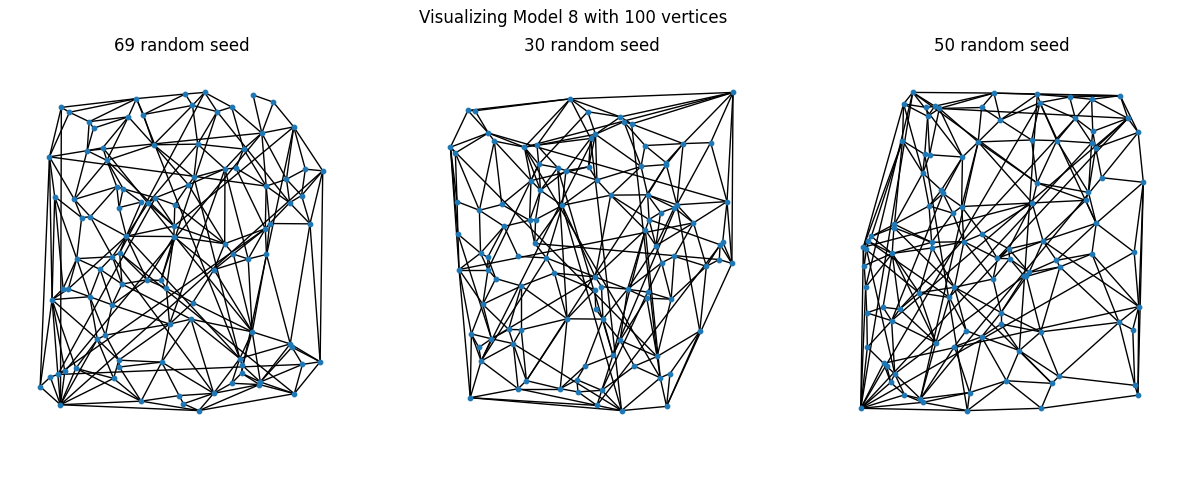}
\caption{Visualizing three different random instances of Model 8 with 100 vertices}
\label{fig:visualize_model8}
\end{center}
\end{figure}

To gain further insight into its structure, we visualized three different random instances of Model Eight. As shown in Figure~\ref{fig:visualize_model8}, the use of preferential attachment results in a more evenly dispersed structure. However, as additional edges are added, some connect vertices that are located on opposite sides of the graph. This reduces the likelihood of planarity and introduces long-range connections that are unlikely to appear in real-world districting maps, where adjacency is typically governed by geographic proximity.

\subsubsection{Model Nine}\label{sec:model9}

Model Nine combines the edge removal strategy of Model Four and the local structure emphasis of Model Seven. Similar to what was observed in Model Four, Model Nine demonstrated that increasing the edge removal probability reduces both the spanning tree constant and the graph’s average degree, potentially distancing the model from real-world properties.

\subsubsection{Model Ten}\label{sec:model10}
Model Ten investigates whether the order in which edges are added and removed affects the average degree and spanning tree constant of the resulting graph. This model inverts the sequence used in Model Nine: it first adds the $n$ shortest edges not yet present in the graph, then removes edges randomly. Two versions of this model were tested, with edge removal probabilities of 0.2 and 0.4. %While structurally identical to Model Nine, Model Ten differs in the order of operations.

As expected, increasing the removal probability decreases both the average degree and the spanning tree constant. However, the ordering of operations proves to be significant. Adding edges prior to removal results in higher values for both metrics compared to Model Nine under the same removal probability, suggesting that initial edge inclusion biases the graph toward greater connectivity before edge pruning occurs.

\subsubsection{Model Twelve}\label{sec:model12}
Since Model Ten demonstrated that the order of adding and removing edges matters, Model Twelve inverts the sequence of Model Eight: first it adds edges using preferential attachment, and then it randomly removes edges. With an edge addition scaling factor of 0.5 and a removal probability of 0.14, this model performed very well.

\begin{table}
    \centering
    \begin{tabular}{|c|c|c|c|c|}
        \hline
        \textbf{Avg. Degree} & \textbf{Median Degree} & \textbf{Max Degree} & \textbf{Avg. ST Constant} & \textbf{ST Asymptote} \\ \hline
        5.45 & 5.39 & 10.5 & 1.44 & $\approx$ 1.6 \\
        \hline
    \end{tabular}
    \caption{Analysis of Model 12}
    \label{tab:model12}
\end{table}

\begin{figure}
\begin{center}
\includegraphics[scale=0.8]{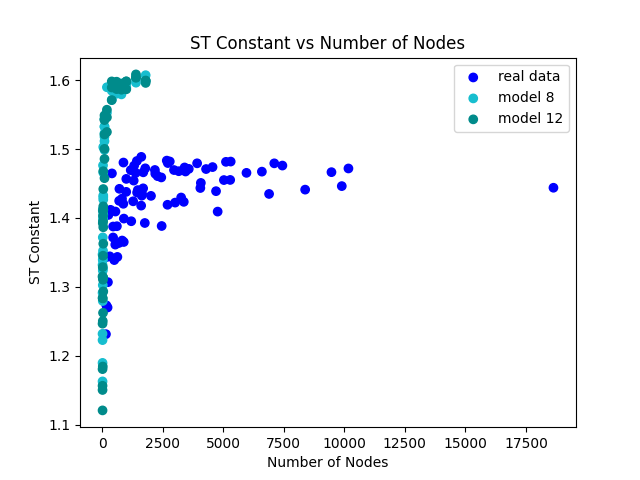}
\caption{Comparing the spanning tree constant for real data to Model 8 and Model 12 results}
\label{fig:st_cons_real_data_m12}
\end{center}
\end{figure}

Figure~\ref{fig:st_cons_real_data_m12} and Table~\ref{tab:model12} show that Model Twelve shares nearly identical characteristics with Model Eight. This suggests that reversing the order of edge addition and removal only affects model characteristics under specific conditions—particularly depending on the nature of the edges being added or removed. The primary difference between the two models is that Model Twelve better approximates the real data, with a marginally lower average degree and average spanning tree constant. However, the asymptotic behavior of the spanning tree constant remains nearly indistinguishable between the two. Given these similarities, Model Twelve provides a more reliable approximation of the real data due to its overall structural consistency.

\begin{figure}
\begin{center}
\includegraphics[scale=0.5, trim = 0 55 0 60, clip]{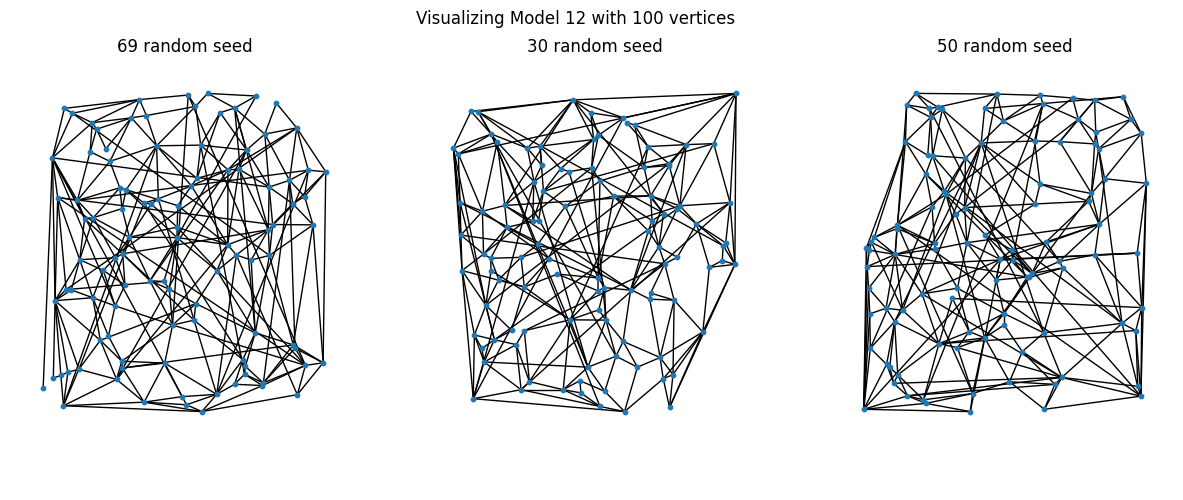}
\caption{Visualizing three different random examples of Model 12 with 100 vertices}
\label{fig:visualize_model12}
\end{center}
\end{figure}

In terms of spatial distribution, Figure~\ref{fig:visualize_model8} (three examples of Model Eight) and Figure~\ref{fig:visualize_model12} Three examples of Model Twelve) highlight that adding edges using preferential attachment leads to a more evenly spread out structure. Nevertheless, in both Model Eight and Model Twelve, as mentioned in Section~\ref{sec:model8}, the process of edge addition occasionally connects vertices located on opposite ends of the graph, which conceptually makes less sense and also makes it much more challenging to maintain planarity.

\subsection{Summary of Random Graph Model Results}\label{sec:RG-sum}

\begin{figure}
\begin{center}
\includegraphics[scale=0.7]{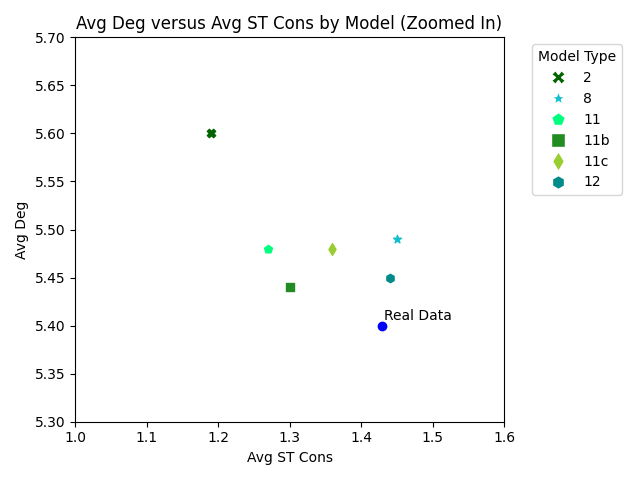}
\caption{Comparing average degree and average spanning tree constant results for the most successful models to these values for the real data}
\label{fig:av_deg_st_cons_by_model_zoomed_in}
\end{center}
\end{figure}

A summary of results for all models is shown in Table~\ref{tab:results}, with the three most successful models (Models 11c, 8, and 12) highlighted. To appropriately compare these models, in Figure~\ref{fig:av_deg_st_cons_by_model_zoomed_in} we plot these models (as well as Models 2, 11, and 11b) according to their average degree and average spanning tree constant to compare to real data.

\begin{table}
    \centering

\begin{tabular}{crrrrrr}
\toprule
\makecell{Model \\ Type} & 
\makecell{Planar} & 
\makecell{Connected} & 
\makecell{Avg. \\ Degree} & 
\makecell{Median \\ Degree} & 
\makecell{Max \\ Degree} & 
\makecell{Avg. ST \\ Constant} \\
\midrule
\rowcolor{CornflowerBlue}
Real Data & 0.72 & 0.89 & 5.42 & 5.00 & 24.28 & 1.43 \\
2 & 0.00 & 1.00 & 5.60 & 5.41 & 11.18 & 1.19 \\
3 & 1.00 & 0.90 & 5.57 & 5.56 & 9.66 & 4.29 \\
4 & 1.00 & 1.00 & 3.60 & 3.64 & 7.49 & 2.46 \\
4b & 1.00 & 1.00 & 2.37 & 2.13 & 5.63 & 1.25 \\
5 & 0.00 & 1.00 & 19.11 & 19.00 & 29.71 & 2.21 \\
5b & 0.07 & 1.00 & 17.73 & 17.60 & 28.03 & 1.95 \\
6 & 0.00 & 1.00 & 7.57 & 7.50 & 13.08 & 6.02 \\
7 & 0.00 & 1.00 & 6.17 & 5.93 & 12.29 & 4.10 \\
\rowcolor{lime}
8 & 0.00 & 1.00 & 5.49 & 5.40 & 10.16 & 1.45 \\
9 & 0.00 & 1.00 & 5.57 & 5.53 & 10.17 & 4.20 \\
9b & 0.02 & 1.00 & 4.08 & 3.93 & 8.31 & 2.73 \\
10 & 0.00 & 1.00 & 6.01 & 5.80 & 11.42 & 4.56 \\
10b & 0.00 & 1.00 & 4.55 & 4.58 & 9.61 & 3.21 \\
11 & 0.00 & 1.00 & 5.48 & 5.27 & 11.73 & 1.27 \\
11b & 0.00 & 1.00 & 5.44 & 5.21 & 11.93 & 1.30 \\
\rowcolor{lime}
11c & 0.00 & 1.00 & 5.48 & 5.25 & 12.20 & 1.36 \\
\rowcolor{lime}
12 & 0.00 & 1.00 & 5.45 & 5.39 & 10.50 & 1.44 \\
\bottomrule
\end{tabular}

\caption{A summary of the empirical properties of all non-grid-based random graph models we considered.  The `Planar' and `Connected' columns refer to the fraction of graphs generated by each model that satisfy those considerations, while the other columns display averages across all random instances we generated for each model. As highlighted, Models Eight, Eleven (c), and Twelve most successfully approximated the quantities of interest from the real data. }\label{tab:results}
\end{table}

\FloatBarrier

\section{Conclusion}
\label{sec:conclusion}

Overall, this paper represents an important first step in exploring the empirical properties of dual graphs. We do not attempt to be comprehensive, but hope that our results can spark further investigation in this area. 

There remain many open questions for further inquiry. Computational limitations prevented more accurate estimations of splittability probabilities in Section~\ref{sec:split}, as well as consideration of $3$-splittability and $4$-splittability for tract and block group data. Another next step is to consider splittability with regard to population, rather than the number of vertices.
We were also limited in only considering our random graph models in Section~\ref{sec:models} on graphs of up to 2500 vertices, and chose not to consider block-level dual graphs in our analysis because of their large size. Moving beyond these computational barriers with different computational approaches (such as using a language that's faster than python) is an obvious next step that could lead to significant additional insights; initial explorations suggest that some of the similarities present in tract and block group graphs may not be present in block graphs. 

There also remain a plethora of other random graph models that could be considered.  What if points are placed in the plane not uniformly at random but according to some other distribution, such as a Gaussian or other clustered distribution?  This could resemble geography more accurately, where there are many small nearby census units in a city but fewer, more spread out census units in rural areas. There are also a variety of additional approaches that could be used to connect these random points, or one could move beyond connecting random point clouds to consider other random graph models. 

Overall, we hope our work can help inform algorithmic research in this area by providing guidance as to the properties of dual graphs encountered in practice. By knowing what to expect for dual graphs, researchers can develop new algorithms that perform better on real-world data. 
They can also rigorously explore the properties of existing algorithms on graph classes -- such as those we suggested as random graph models -- that more closely resemble those encountered in practice.  We hope our work is just the first step in this important new direction.

\section{Further Materials}

A large portion of the work done in this paper was originally part of Brooke Feinberg and Anne Friedman's Senior Theses at Scripps College during the 2024-2025 academic year~\cite{bf-thesis, af-thesis}, both advised by Sarah Cannon. 

Code for the experiments and visualizations presented in this paper can be found at: 
\begin{itemize}
    \item \url{https://github.com/sranders15/Spanning-Trees-of-Dual-Graphs} for some work presented in Section~\ref{sec:props} and all work in Section~\ref{sec:grids}. Some of this code is heavily based on work in~\url{https://github.com/brookecarofeinberg/SamplingBalancedForests}, though a small bug present in the latter has been corrected. 
    \item \url{https://github.com/afr13dman/senior-thesis} for some work presented in Section~\ref{sec:props} and all work presented in Sections~\ref{sec:RGs} through~\ref{sec:RG-sum}. 
    
\end{itemize}

%\end{itemize}

%\nocite{*} 
%\bibliographystyle{plain}
%\bibliography{citations}

\end{document}